# Remnant magnetization above room temperature in semiconducting $Y_{0.5}Ca_{0.5}BaCo_4O_7$

Martin Valldor

Institut für Physik der Kondensierten Materie, TU-Braunschweig, Mendelssohnstrasse 3, D-38106 Braunschweig, Germany
E-mail: m.valldor@tu-bs.de

## Abstract

The $Y_{0.5}Ca_{0.5}BaCo_4O_7$ compound exhibits four magnetic anomalies at the temperatures 387, 281, 52, and 14 K; all anomalies show characteristics typical for spin freezing into disordered states: frequency dependent transition temperature in the *AC* magnetic susceptibility together with relaxation of thermoremnant magnetization. $Y_{0.5}Ca_{0.5}BaCo_4O_7$ is a semiconductor with a small band-gap of 0.17 eV concluded from four-point conductivity measurements and the conductivity is proportional to $T^{-3/4}$, suggesting an electron hopping mechanism involving $Co^{2+}$ and $Co^{3+}$ ions; at higher temperatures, double-exchange is the proposed reason for the strong magnetic interaction. At lower temperatures, super-exchange interactions start to compete with the double-exchange for domination and this competition together with geometrical frustrations in the structure are responsible for magnetic disorder down to 2 K. Covalence between Co and O is also discussed as additional reason for the strong magnetic interactions.

## INTRODUCTION

Transition metal oxides (TMOs) have been investigated more intensively the last decades, due to the many new properties found in this area *e.g.* superconductivity and colossal-magneto resistivity. Lately, TMO compounds with interesting complex properties have been discovered, of which diluted magnetic semi-conductors [1] and magnetic ferroelectrics [2] are two with promising applicability. The strong electric and magnetic correlations in these TMOs have partly been understood and among the commonly accepted magnetic TM-TM interaction mechanisms are: direct-exchange, super-exchange [3], double-exchange [4], and RKKY [5]. At present, most of the investigations are focused around the perovskite structure, often denoted $ATMO_3$ ($A$ = rare-earth- or alkali-earth metal), since it is relatively simple *i.e.* the symmetry is high, the unit cell is small and the TM-O-TM orbital overlap is close to ideal for magnetic interactions. However, interesting magnetic behaviors can also be found within other structure types *e.g.* spinel [6], pyrochlore [7], corundum [8], and magnetoplumbite [9]. In these four examples, geometrical frustrations, meaning triangular or tetrahedral units of anti-parallel magnetic spins, are found and the properties in these systems are described as either disordered anti-ferromagnets, spin-glasses, or cluster formations of magnetic units *e.g.* super-paramagnets.

In an earlier paper, the compound $YBaCo_4O_7$ [10] was reported and its new Co-O substructure contains a geometrical frustration; the large frustration was evident, because the Curie-Weiss constant (-907 K) was much larger than the actual transition temperature (65 K), and the magnetic properties were described as complex with features similar to spin-glass materials [11]. As there are numerous papers where large changes in electric and magnetic properties are observed as function of mean valence of the TM ion [12], it was tried to replace half of Y by Ca, *i.e.* changing the mean valence state of Co, in $YBaCo_4O_7$ to see the effects on the frustration and the magnetic properties. The results of this investigation are presented below.

## EXPERIMENTAL

The sample was prepared through solid-state reactions of $Y_2O_3$ (Alfa 99.99%), $CaCO_3$ (Merck >99%), $BaCO_3$ (Merck *p.a.*), and $Co_3O_4$ (Baker Chem. *p.a.*). Stoichiometric amounts were weighed in and mixed in an agate mortar before reacting the powder during 10 hours in air using a corundum crucible and a temperature of 1100 °C. The sintered sample was reground a fired again at the same temperature over night. The cooling after the second heating step was rapid; the sample was cooled inside the furnace down to 900°C and subsequently quenched to room temperature.

Cell parameters were refined from powder X-ray diffraction data obtained with a Guinier focusing camera (radius = 40 mm, $CuK\alpha_1$ - $\lambda$ = 1.540598 Å), using pure $SiO_2$ as an internal standard. An image plate, scanned with a BAS-1800 II scanner from FUJIFILM, was used as detector system. Further powder diffraction data, used in Rietveld refinements, were collected with a STOE STADI-P diffractometer in Debye-Scherrer transmission mode using $CuK\alpha_1$ radiation ($\lambda$ = 1.540598 Å). The investigated diffraction range was 15–100° (2θ) and the data were modeled using Fullprof2000 [13]. The refinement was performed as follows: the background was set manually, the Pseudo-Voigt function was used to model the peak shape, and peak asymmetry was accounted for up to 2θ = 50°. All positions were assumed to be fully occupied and the Ca:Y ratio was set to 1. The refined parameters were: scaling factor, zero-position, cell parameters, 4 variables for the peak shape, 2 peak asymmetry parameters, 9 atomic positional parameters, 5 isotropic displacement parameters (the oxygen atoms were refined as one).

Scanning electron microscope (SEM) studies together with energy dispersive spectrometry (EDS) analysis were made in a Leica Stereoscan 420i at 15 kV equipped with an Oxford Instrument INCA V.4.01. Ten different crystallites were checked for composition.

Magnetic susceptibility measurements in the range 2–350 K were performed with PPMS from Quantum Design; the *AC* measurements were made with 125–9975 Hz of a 10 Oe field. *DC* magnetic measurements were made with a SQUID MPMS (Quantum Design) in the range 5-400 K using fields of up to 5 T. The high-temperature range (300-700 K) was realized in a SQUID MPMS-S5 (Quantum Design) with a built-in furnace. Complementary high-temperature magnetic measurements in the range 300-500 K were done using a VSM from Oxford Instruments with a field of 0.5 T.

The four-point conductivity measurement was made between 110 and 300 K on a sintered pellet and the contacts were fastened with silverpaint. The voltage was kept constant at 2 mV using a resistance bridge from Linear Research Inc. and the measuring range was up to 2 MOhm.

## RESULTS

*X-ray analyses*

The powder sample proved to be X-ray pure according to the Rietveld pattern (Fig.1), where only the desired phase was observed. The relatively high background intensity, due to the strong presence of Co, caused the conventional *R*-values to be larger than expected, however the $\chi^2$ and the GoF were both satisfactory (Table I).

The elemental analyses (Table I) showed that the compound is homogeneous, however a slight overpresence of Y was detected. Y ($Y^{3+}_{radius}$ = 1.04 Å) is too small for the Ba site ($Ba^{2+}_{radius}$ = 1.75 Å) and too large for the Co site ($Co^{2-3+}_{radius} \approx 0.72$ Å [15]) giving only one possible explanation: the Y standard (Y-metal) was not ideal for the here presented oxide.

*Structure description*

To better understand the magnetic properties, the Co and the Co-O substructures have to be described. In Fig.2(a), the Co2 sites form *kagome*-like layers that are connected via the apical Co1 sites. The Co-O substructure in Fig.2(b) is built up merely from corner-sharing $CoO_4$-tetrahedra: Co1 is found in the relatively smaller tetrahedron that connects the planes of the relatively larger Co2 tetrahedra. The symmetrical arrangement of tetrahedra results in a geometrical frustration and all Co-O-Co bond angles are close to the tetrahedral angle (109.5°). The smaller Co1 site should stabilize a higher valence than Co2 and according to a bond-valence-sum (BVS) calculation [16], the valency at the sites are 2.525 and 2.124, respectively.

The Co-O substructure can also be described as a hollow wurtzite type structure, however all tetrahedra are somewhat distorted; "hollow" while parts of the tetrahedra in wurtzite are have been replaced by Ba. The structure cannot hold oxygen defects, other than Schottky defects, since all oxygen atoms are bound within the tetrahedral net and a missing oxygen atom would cause Co to be trigonal planar coordinated, which is highly unlikely considering the space and the coordinational environment. This restriction on the oxygen content is favorable in this case, since the Co mean oxidation state can easily be calculated; the stoichiometry can be written $Y_{0.5}Ca_{0.5}Ba(Co^{2+})_{2.5}(Co^{3+})_{1.5}O_7$ and the Co ions have to be distributed on three Co2 and one Co1 sites.

The two possible oxidation states of Co, in this structure, are +2 and +3; the electronic configuration of $Co^{2+}$ is $d^7$ and that of $Co^{3+}$ is $d^6$. When adding two known facts together, *i.e.* (i) oxygen normally induces a relatively small crystal-field splitting of $t_2$ and $e$ states and (ii) the tetrahedral field splitting is less than half (4/9) of the octahedral one, the high-spin state of Co seems more likely; this means that $Co^{2+}$ is a $S$ = 3/2 and $Co^{3+}$ is a $S$ = 2 ion. The symmetry of the Co1 site is *3m* and for Co2 only *m*; the three-fold symmetry at the Co1 site causes a severe conflict with the possible orbital configurations, where an orthogonal setting is expected. At the Co2 site, no direct symmetry problem is encountered, however three Co2 tetrahedra meet at the three-fold axis, inducing geometrical frustration.

*Magnetic properties – AC susceptibility*

(a) *Composition*. As can be seen in Fig.3, the presence of Ca, *i.e.* the $Co^{2+}/Co^{3+}$ ratio, has a great impact on the magnetic properties. The compound without Ca [10] exhibits one obvious transition at about 65 K and most likely a second starting close to 5 K [11], but $Y_{0.5}Ca_{0.5}BaCo_4O_7$ has three clearly visible "transitions", from here on called the freezing temperatures ($T_f$s), in the range 2-350 K: at 281, 52, and 14 K. The first $T_f$ starts already close to 300 K, where an increase in $\chi$ is seen. Below the broad transition close to 280 K, $\chi$ decreases until almost 200 K. Also at 14 K, a broad transition is seen, but the transition at 52 K is relatively sharp.

(b) *Frequency dependence*. Using several different frequencies of *AC*-susceptibility, it was possible to see frequency dependence at the transitions; the data for the transition close to 281 K are shown in Fig.4. Similar behaviors were observed for the other two transitions at 52 and 14 K. The frequency dependent transition temperatures ($T_f(\omega)$) are further discussed and modeled below.

The imaginary parts of the *AC*-susceptibility are presented in Fig.5. It was postulated by Casimir-duPre [17] that, close to magnetic transitions, the average spin relaxation time ($\tau_{AV}$) is inverse proportional to the measuring frequency ($\omega$) causing the largest imaginary contribution ($\chi''_{max}$). Hence, for the transitions at 281 and 52 K, where the 9975 Hz frequency measurement has the highest $\chi''$, is the average relaxation rate shorter than 0.1 ms. However, for the transition at 14 K, the measurement at 4025 Hz gives the highest $\chi''$, which is a clear indication of a change in relaxation rates; $\tau_{AV}$ has drastically increased to about 0.25 ms.

Note, that the imaginary part, with respect to different temperatures, is not on the same level even at 350 K and this initiated the search for a further magnetic transition at higher temperatures, as all imaginary parts should be zero and equal for all frequencies in the paramagnetic range.

(c) *Harmonics of $\chi$*. The different harmonics of $\chi'$ and $\chi''$ are displayed in Fig.6. The first harmonic of the real part susceptibility ($\chi'_I$) exhibits three possible magnetic transitions, also mentioned above, at 281, 52, and 14 K, respectively. For $\chi'_{II}$, there is only one strong contribution at 281 K. The imaginary parts complex matters: $\chi''_I$ also contains three features (277, 50.5, and 11 K), of which all appears just below the respective $\chi'_I$. Two deviations from zero are apparent for $\chi''_{II}$ at 280.7 and 69.7 K, of which the first is negative and the second positive. Only very small intensities can be seen in the third harmonics data *e.g.* a small contribution in $\chi'_{III}$ at about 280 K that makes it even more obvious that this transition lacks spontaneous character.

Due to the strong negative contribution in $\chi''_{II}$, it is obvious that the transition around 280 K is a complex of at least two competing interaction mechanisms; two parts can be separated by the different harmonics of $\chi''$: a spontaneous magnetic transition starts at 280.7 K, but the magnetic structure loses energy at 277 K; either parts of the magnetic order are lost due to a second competing magnetic interaction or the magnetic domains are very small and the domain boundaries represent a large fraction of the total magnetic volume, where aligning problems obviously occur, leading to a domain-growth problem.

At about 70 K, a broad positive peak in $\chi''_{II}$ signals the oncoming of a new energy minimum for the magnetic structure, but again this magnetic energy is lost at 50.5 K and the progression towards long-range magnetic order is once more hindered.

(d) *Modeling of $T_f(\omega)$*. The relation between the magnetic freezing temperature ($T_f$) and measuring frequency ($\omega$) has been investigated for several other magnetically disordered systems (see for example [18,19]). For clear domain formations, *e.g.* super-paramagnetism, the Arrhenius plot approach is enough to extract reasonable activation energies. In the present case ($Y_{0.5}Ca_{0.5}BaCo_4O_7$), this fails utterly (see Table II) for all three transitions, suggesting that the frozen magnetic states are other than super-paramagnetic. The Vogel-Fulcher [20] and the power-law models have also been used in other works to extract information about non-spontaneous magnetic transitions. That $Y_{0.5}Ca_{0.5}BaCo_4O_7$ data are plotted in Fig.7(a,b) and show relatively good fitting, especially for the two transitions at higher temperatures. The obtained values, from the linear curve fits, are presented in Table II. The activation energy ($E_a$) is obtained from the Vogel-Fulcher model and can be interpreted as the energy to start decoupling the magnetic spins and is very low for all transitions, meaning high degeneracy of the magnetic states. The dynamic exponents ($zv$) shown in Table II are for all three transitions well within the expected values for spin-glasses *i.e.* $zv = 5–12$, although other measurements (see below) clearly indicate that the compound is not at spin-glass. The reason for this ambiguous behavior is, up to now, not fully understood. The $\omega_0$ values, obtained from the power-law fitting, seem reasonable when comparing to what is usually seen for similar systems.

*Magnetic properties – DC susceptibility*

(a) *FC-ZFC*. In the graphs shown in Fig.8 there is an obvious difference between field-cooled (FC) and zero-field-cooled (ZFC) data. Already at 450 K, a faint difference is present, but the first visible transition-like turn appears at 387 K. All four "transitions", visible in the ZFC data, are smeared out in the FC data, but changes in the FC slope appear at about the same temperatures.

The ZFC data with and without furnace in Fig.8 superimpose well, however the low-temperature FC data should probably be shifted to fit the high-temperature data.

From the dashed line in Fig.8, it was possible to calculate the magnetic moment per Co, by using the Curie-Weiss formula: 6.88 $\mu_B$. This is much higher than expected for $Co^{2-3+}$ ($Co^{2+}$: $\mu_{eff.} = 3.87$, $\mu_{exp.} = 4.3–5.2$ $\mu_B$, $Co^{3+}$: $\mu_{eff.} = 4.9$, $\mu_{exp.} = 4.3$ $\mu_B$ [20]) and could be due to electron conductivity or strong magnetic fluctuating interactions in the temperature range used for the calculation. The large negative Weiss constant ($\theta_{CW}$ = -2203 K) points at extremely strong magnetic interactions.

To confirm the highest transition temperature, an additional measurement was performed in a VSM (data not shown here): the difference between FC and ZFC data below ~390 K was obvious and both curves superimpose above this temperature.

(b) *Magnetization*. Isothermal magnetization was performed and the results are displayed in Fig.9. At all investigated temperatures, the magnetization is small and very similar for a wide range of temperatures. There is only one explanation for this behavior: at all temperatures below the first transition the resulting magnetic moment is low. There is no tendency for reaching saturation at 5 T for any temperature, indicating strong interactions without any meta-magnetic contributions.

There is a small isothermal magnetic remanence at 5 and 40 K indicating a faint hysteresis at lower temperatures.

(c) *TRM and aging*. Relaxation of thermoremnant magnetization (TRM) was observed at all measured temperatures below 387 K and the data obtained at 370 K is displayed in Fig.10. Aging does not affect the frozen magnetic state, since the three different waiting times result in similar relaxation processes, which means that $Y_{0.5}Ca_{0.5}BaCo_4O_7$ is not a spin-glass. Instead of a purely logarithmic decay of TRM, which cannot fully describe the system at hand, the relaxation is more similar to a logarithmic function with a superimposed stretched exponential function [22], proven by the fact that the data with the aging time $10^3$ s have a change of slope around $10^3$ s, but the $10^4$ data have not.

The relaxation of TRM at all temperature below 387 K suggests that the magnetic systems is disordered in it frozen state and the reason for this behavior will be discussed below.

*Conductivity investigations*

The electronic resistivity ($\rho$) shows the typical behavior of a semiconductor with a relatively small gap and the gap size was estimated to 0.17 eV (Fig.11). The only magnetic transition present in the temperature range, at 280 K, does not obviously change the curvature and is therefore not connected to a charge ordering effect. If the inset in Fig.11 is carefully examined, it is possible to see that the straight line cannot fully describe the observed $\ln\rho(T^{-1})$ data. Plotting the resistivity data as function of $T^{-1/4}$ does not give a straight line either, indicating that the moving charges cannot be described with the Motts 3D variable range hopping model for disordered systems [23]. It was proposed that the $\ln(\rho)$ should be proportional to $T^{-1/2}$ if the charge carriers strongly interact with eachother in a hopping model [24]. Furthermore, even the concentration of the charge carriers has been discussed to play an important role in the size of the exponent [25]. The data from $Y_{0.5}Ca_{0.5}BaCo_4O_7$ is best fitted with $T^{-3/4}$, meaning that the conductivity is not only thermally activated and this could be due to high concentration of charge carriers and/or the fact that the dimensionality of the conductivity is low, as will be discussed below.

DISCUSSION

$Y_{0.5}Ca_{0.5}BaCo_4O_7$ contains a geometrical frustration around three-fold symmetry axes and mixed valency ($Co^{2-3+}$) is found at two sites in the structure. This is also the case in $YBaCo_4O_7$ [10], but the magnetic properties are very different, due to the small change in the mean Co valence. To be sure that the structure is of less importance than the charge change, when identifying the strong magnetic interactions, the $YBaCo_4O_7$ structure is compared with the one presented in this paper and the bond-angles are presented in Table III. $YBaCo_4O_7$ has a transition to a disordered anti-ferromagnetic state at 65 K, a $\theta_{CW}$ = -907 K, and an average oxidation state of Co = +2.25. The structure refinement of $Y_{0.5}Ca_{0.5}BaCo_4O_7$ presented above was performed at room-temperature (c:a 95 K below $T_f$) and the corresponding angles are also presented in Table III; the average Co oxidation state is +2.375 and the $\theta_{CW}$ = -2203 K. Most of the bond angles are close to the ideal tetrahedral angle, however the variations closing up on 90° should result in relatively stronger super-exchange [3] interactions. The only hints that a change in structure between $Y_{0.5}Ca_{0.5}BaCo_4O_7$ and $YBaCo_4O_7$ would affect the properties is that the Co2-O1-Co2 bond angle should be large and the Co2-O3-Co1 bond angle small, but this cannot not alone be the reason for such a large change in properties, since the angle changes are small. Therefore, the charge change is the only reasonable explanation for the stronger magnetic interactions in $Y_{0.5}Ca_{0.5}BaCo_4O_7$ compared to $YBaCo_4O_7$.

Strong anisotropy is also expected in the system; the double-exchange should only occur along the unique axis, since any other in-plane exchange would result in metallic conductivity, as all the in-plane sites are symmety equivalent and the band-gap in that case had to be zero. These kinds of magnetic "columns" have also been used to explain the magnetism of the $BaCoO_3$ [26] and $Ca_3Co_2O_6$ [27].

The low magnetic moment observed in the magnetisation measurements (Fig.9) needs to be discussed; it has been observed that especially Co, in octahedral oxygen coordination, can experience spin-moment quenching by spin-orbit coupling (see for example [28,29]). Similar examples are lacking for the tetrahedrally coordinated Co in oxides, however it is likely that the same quenching could also occur in this case. If a low-spin tetrahedral coordination could be realized, the observations could be explained, but such phenomena has up to now only been seen for metalorganic compounds [30,31] and further investigations have to be done to clear up this point.

There is, however, one further possibility: the system is more covalent. Covalency has been found for Co in a MgO matrix studied by optical means [32], but calculations of the same coordination (octahedral) for Co ($CoO_6^{10-}$) reveals no covalent bonding character [33]. However, recent observations [34] of the $LaCoO_3$ perovskite show that strong covalent character is present between Co and O and in the surface between Co(metal) and $Al_2O_3$ the bonding exhibits covalent nature, where electrons are partly given to the oxygen and the magnetic moment of Co is lowered [35]. Hence, all oxygen atoms in $Y_{0.5}Ca_{0.5}BaCo_4O_7$ are assumed to be hybridized (O1+O3 $sp^2$ and O2 $sp^3$), as in a covalent situation. This would explain the strong magnetic interactions, since 120° or the tetrahedral angle, respectively, would have the largest orbital overlap and it could also explain the relatively low magnetisation observed in the *DC* field (Fig.9), as the electron would stay longer on the oxygen atoms (in the $sp^2$ or $sp^3$ orbitals), where no or only low orbital momentum is present. Even more important, as proof of hybridization of the oxygen orbitals, is the fact that the hybridizations does not break any crystallographic symmetries, as non-hybridized oxygen orbitals do, meaning that the hybridizations are advantageous situations.

Although relaxation of TRM is observed (Fig.10), the relaxation process at 370 K must be clarified. As the double-exchange is strong along the unique axis (*c*) and does not connect in the planes, the magnetic spins will form columns along *c*, which will be quasi-one-dimensional. In one column, the spins are more or less aligned due to the spin-direction preservation *i.e.* the mediating electrons do not change their spin-direction. However, the spin directions among the columns can be random (parallel or anti-parallel), since the inter-column interactions at this temperature are weak. This will lead to a disorder of the "spin-column" directions and in the ZFC case, an almost statisitical distribution of directions will be the result, only affected by the weak inter-column couplings. In the FC situation, more of the spin-columns are forced to align along the outer field, giving a higher magnetisation. When the outer field is shut off after FC, the weak inter-column interactions start to redirect the columns to reach randomness, which is thermodynamically more stable, as the entropy is increased. It is here worth mentioning that the magnetic columns constitute a triangular lattice; the inter-columnar interactions are frustrated. A similar columnar behavior was also discussed for $Ca_3Co_2O_6$ [27].

There are seldom more than two magnetic "transition" for a pure sample, but in this case it seems as if all magnetic observations belong to the same phase, $Y_{0.5}Ca_{0.5}BaCo_4O_7$. As comparisons, CoO turns antiferromagnetic at $T_N$ = 292 K, $Co_3O_4$ at $T_N$ = 40 K and $BaCoO_{3-\delta}$ at $T_N$ = 8 K, of which all are spontaneous transitions *i.e.* do not show frequency dependent *AC* susceptibility or relaxation of TRM. $BaCoO_{3-\delta}$ has also magnetic domains forming at 250 K [26] and below 40 K a gap opens up between ZFC and FC magnetizations [36], indicating magnetic degeneracy. $YBaCo_2O_5$ is also known, but it turns into an ordered antiferromagnetic state (G-type) at $T_N$ = 330 K [37]. As conclusion, none of the magnetic properties from the just mentioned compounds fit the data presented here. Hence, $Y_{0.5}Ca_{0.5}BaCo_4O_7$ is alone responsible for the extraordinary magnetic properties.

Thus, the properties of $Y_{0.5}Ca_{0.5}BaCo_4O_7$ can tentatively be described as follows: along the unique axis (*c*) there is a strong coupling setting in at 387 K, described as double-exchange between $Co^{2+}$ and $Co^{3+}$, with a low activation depending on the small difference in crystal field energy between two sites in the structure. Super-exchange couplings build up a geometrically frustrated substructure in the planes of Co2 (Fig.2a) and, as the temperature is lowered, the combination between super-exchange and double-exchange causes the magnetic system to degenerate. On further cooling, the strength of the double-exchange diminishes, since the electrons do not have enough energy (heat) to hop between the Co atoms. At the same time, the super-exchange starts to couple in more directions and, close to 52 K, the in-plane interactions, of frustrated anti-parallel nature, dominate and cause the magnetic susceptibility to decrease. At 14 K, there seems to be a re-ordering of spins but also this transition lacks spontaneity and short-range magnetic order is most probably the result. Hence, the overall magnetic interactions can be described as those of a glassy disordered anti-ferromagnet, where double-exchange dominates at higher temperatures and super-exchange at lower and they compete through the whole temperature range on a frustrated lattice. Note, that there are other systems exhibiting similar properties, but the magnetic order is in those cases confined within planes, prohibiting long-range order to occur [38].

Thus, there are still several point that need more attention and further investigations of this phase are already planned in the near future.

ACKNOWLEDGEMENTS

I would like to thank Rainer Pöttgen and Rolf-Dieter Hoffman for letting me use their PPMS at Münster University. Peter Lemmens and Bernard Chevalier are acknowledged for their helpful comments on magnetism. I thank Heiko Ahlers and Joachim Lüdke for giving me the chance to use the SQUID at PTB (Braunschweig). Christopher Mennerich is acknowledged for helping me with the VSM and Matthias Bleckmann for introducing me to the resistivity equipment.

Table I. Obtained data from Rietveld refinement. One STD is presented within the parentheses and these values are already multiplied by the Bérar-Lelann factor [14]. The data were obtained at 295 K.

| Compound | | $Y_{0.5}Ca_{0.5}BaCo_4O_7$ | |
|---|---|---|---|
| EDS results | | $Y_{0.61(4)}Ca_{0.50(3)}Ba_{1.07(3)}Co_{3.82(4)}O_x$ | |
| *Position* | Atom Occ. | x, y, z | $Beta_{iso}$ (Å$^2$) |
| 2b | Y 0.5 Ca 0.5 | ⅔, ⅓, 0.8719(8) | 3.1(3) |
| 2b | Ba 1.0 | ⅔, ⅓, ½ | 2.3(1) |
| 2a | Co 1.0 | 0, 0, 0.440(2) | 1.9(2) |
| 6c | Co 1.0 | 0.1704(7), 0.8296(7), 0.687(1) | 2.3(1) |
| 6c | O 1.0 | 0.474(3), 0.526(4), 0.765(4) | 5.9(5) |
| 2a | O 1.0 | 0, 0, 0.252(6) | 5.9(5) |
| 6c | O 1.0 | 0.163(3), 0.837(3), 0.486(4) | 5.9(5) |
| Symmetry | | $P6_3mc$ | |
| a, c (Å) | | 6.30932(3), 10.24912(5) | |
| $\chi^2$ | | 1.31 | |
| $R_{Bragg}$ (%) | | 11.1 | |
| $R_F$ (%) | | 11.8 | |
| GoF ($R_{wp}/R_{exp}$) | | 1.10 | |
| Param./data | | 14(24)/99 | |
| Guinier a, c (Å) | | 6.302(3), 10.240(2) | |

Table II. The calculated values form the least-square fittings in Fig.7(a,b).

| Equation | Arrhenius | | Vogel-Fulcher: y = a×(x-b) | | Power-law: y = d×x+e | | |
|---|---|---|---|---|---|---|---|
| Approx. $T_f$ | $E_a$ [K] | $\omega_0$ [Hz] | a($E_a$)[K] | b($T_0$) [K] | d(zv) | e | $\omega_0$ ($e^e$) [Hz] |
| 280 K | $2\times10^5$ | $1\times10^{323}$ | 0.382 | 278 | 8.6 | 45.8 | $7.8\times10^{19}$ |
| 52 K | $10^4$ | $1\times10^{94}$ | 0.249 | 50.2 | 7.1 | 30.9 | $2.6\times10^{13}$ |
| 14 K | 372 | $7\times10^{14}$ | 0.485 | 10.3 | 6.4 | 14.1 | $1.3\times10^6$ |

Table III. Co-O-Co bond angles inside the tetrahedral net.

| Compound | $Y_{0.5}Ca_{0.5}BaCo_4O_7$ | $YBaCo_4O_7$ [10] | $YBaCo_4O_7$ [10] |
|---|---|---|---|
| Temperature (K) | 295 (95 below $T_f$) | 295 | 10 K (55 below $T_f$) |
| Co2-O1-Co2 (plane) | 113(1)° | 106.5(2)° | 114.9(7)° |
| Co2-O2-Co2 (plane) | 109.3(1)° | 109.2(2)° | 107.4(7)° |
| Co2-O2-Co1 (apical) | 109.7(5)° | 109.7(2)° | 111.4(7)° |
| Co2-O3-Co1 (apical) | 107(1)° | 111.6(3)° | 101.2(6)° |

Figure Captions

Fig.1 Rietveld pattern from the structure refinement of $Y_{0.5}Ca_{0.5}BaCo_4O_7$. Observed data are marked with circles and the full line represents the calculated pattern. The Bragg positions are indicated with vertical lines and below them is the $Int_{obs}-Int_{calc}$ plotted.

Fig.2 (a) The Co substructure of $Y_{0.5}Ca_{0.5}BaCo_4O_7$. Dashed lines are added to outline the network in. (b) A polygon version of the Co-O substructure, where the O atoms are marked with small balls at the corners of the polyhedra. The unit cell is marked with a thin line and the two respective Co sites are indicated with Co1 (light grey) and Co2 (dark grey).

Fig.3 Magnetic susceptibility measured with and *AC* field (10 Oe, 1000 Hz) for $Y_{0.5}Ca_{0.5}BaCo_4O_7$. The dashed line represents the data of $YBaCo_4O_7$ [10].

Fig.4 The magnetic susceptibility (1$^{st}$ harmonic, real part - χ´) for $Y_{0.5}Ca_{0.5}BaCo_4O_7$, using an *AC* field of 10 Oe, plotted *vs*. temperature. The measuring frequencies are indicated according: 325 (⊕), 525 (■), 825 (△), 1000 (□), 2025 (+), 4025 (●), 6525 (⊠), 9975 (O) Hz. The lines are refined 3$^{rd}$ degree equations to fit the data in the range 274-286 K when extracting maxima.

Fig.5 The magnetic susceptibility (imaginary part - χ´´) for $Y_{0.5}Ca_{0.5}BaCo_4O_7$, using an *AC* field of 10 Oe, plotted *vs*. temperature. The individual curves correspond to different measuring frequencies (125–9975 Hz). The data hast been noice-filtered, by averaging over 5 points, to make the figure clearer.

Fig.6 1$^{st}$ and 2$^{nd}$ harmonics of χ (real part – a, imaginary part – b) for $Y_{0.5}Ca_{0.5}BaCo_4O_7$ using a field of 10 Oe at 1000 Hz. Arrows indicate the axis for the respective data. The 1$^{st}$ harmonics are marked with lines and the 2$^{nd}$ with dots.

Fig.7 Vogel-Fulcher plot (a) and power-law plot (b) of the $T_f(\omega)$ data of χ´ (1$^{st}$ harmonics) from Fig.6. The data are presented according to their approximate $T_f$s: 280 K (O), 52 K (▽), and 14 (□). The straight lines are least-square fits and the results of the fits are presented in Table I.

Fig.8 Field-cooled (FC) and zero-field-cooled (ZFC) magnetic susceptibility (a, χ) and inverse mangetic suseptibility (b, $χ^{-1}$) data for $Y_{0.5}Ca_{0.5}BaCo_4O_7$ using a field of 1 kOe (0.1 T). The full lines are made in the SQUID without furnace (0–400 K) and the dotted lines with furnace (300–700 K). The dashed line in (b) represents the extension of the range 600–700 K to extract the Curie-Weiss constants.

Fig.9 Magnetisation plotted *vs*. applied field (Oe) for $Y_{0.5}Ca_{0.5}BaCo_4O_7$. The temperatures are indicated close to the corresponding curve. All of the curves contain data for increasing and decreasing fields.

Fig.10 The relaxation of TRM for $Y_{0.5}Ca_{0.5}BaCo_4O_7$ plotted as function of time. $M_0$ is the first measurement made before the applied field was shut off (at zero time). $M_T$ is the remanent magnetisation in the sample at the time indicated on the x-axis. The different waiting times at 370 K are indicated with different symbols according to the upper right inset. Measuring procedure: start at 450 K – 1000 Oe applied – cool to 370 K – waiting time(indicated in the upper right inset) – $M_0$ measured – field off – start measurement time at zero. The inset (low left) contains the same data but the time scale is linear.

Fig.11 Specific resitivity ($ρ_S$) plotted as function of temperature. The inset contains the same data, but with logarithmic $ρ_S$ data *vs*. inverse temperature. The dashed line is a linear least-square fit.

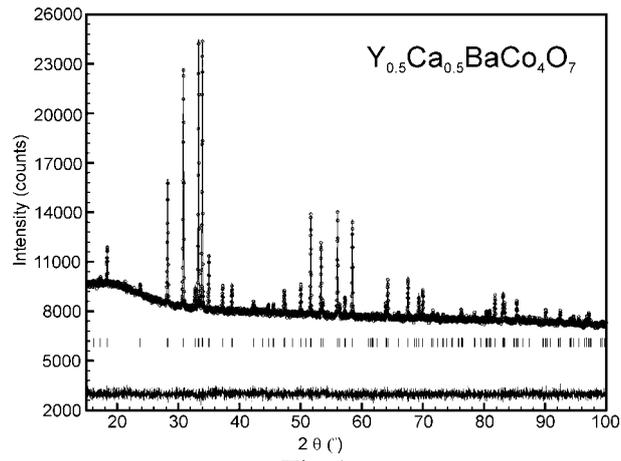

Fig.1

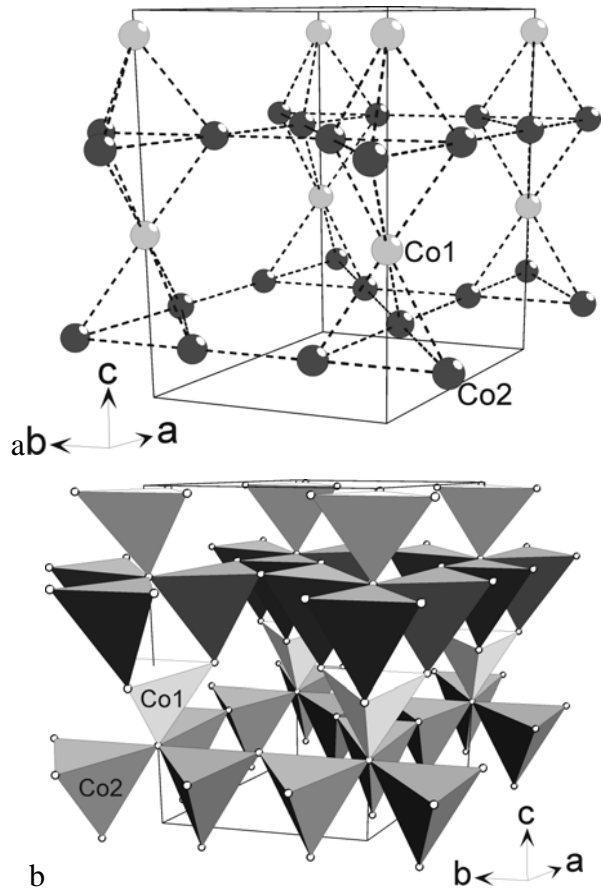

Fig.2

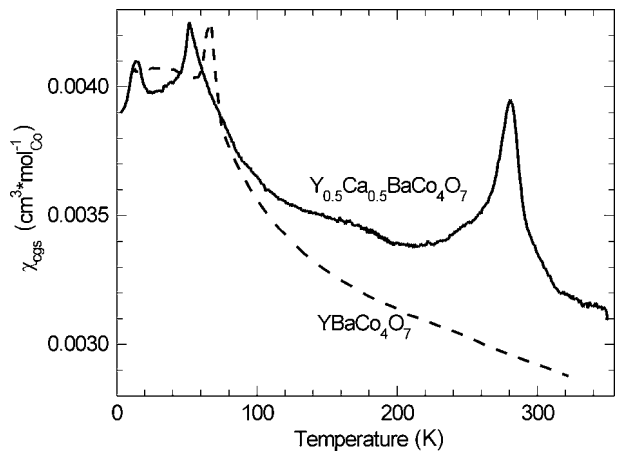

Fig.3

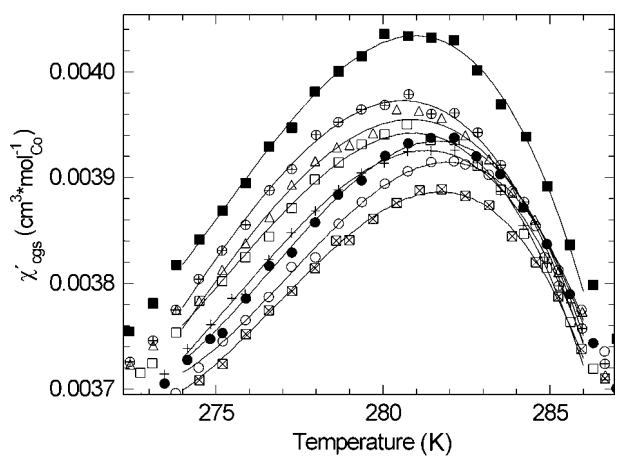

Fig.4

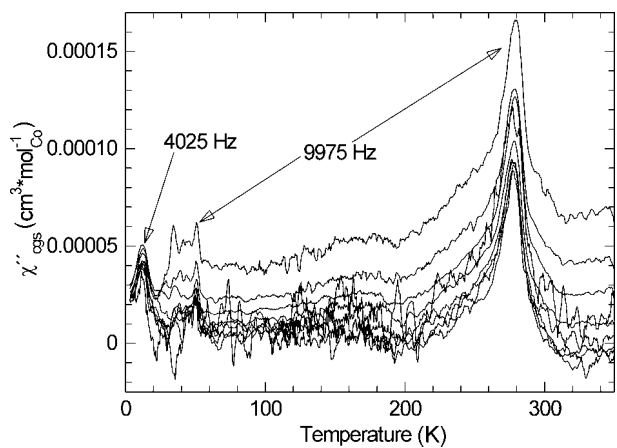

Fig.5

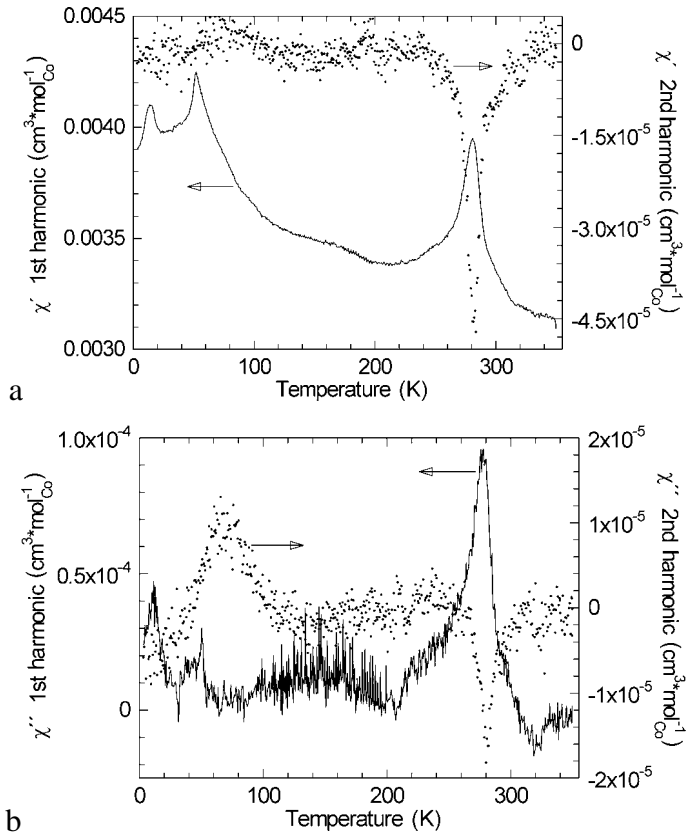

Fig.6

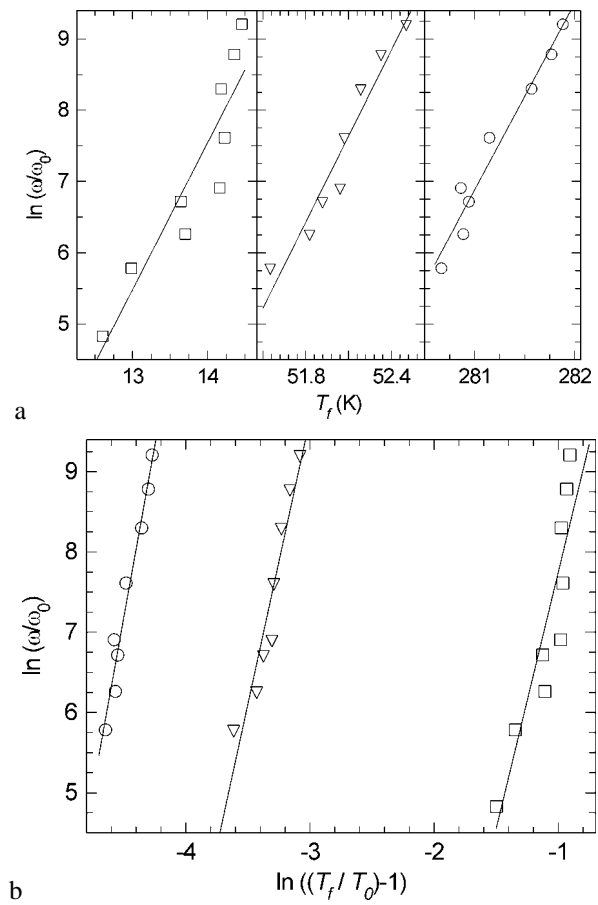

Fig.7

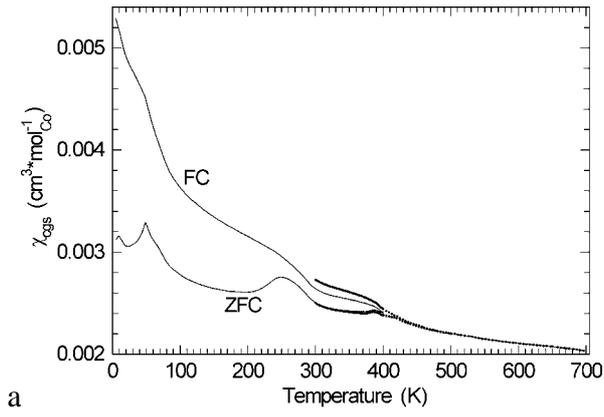

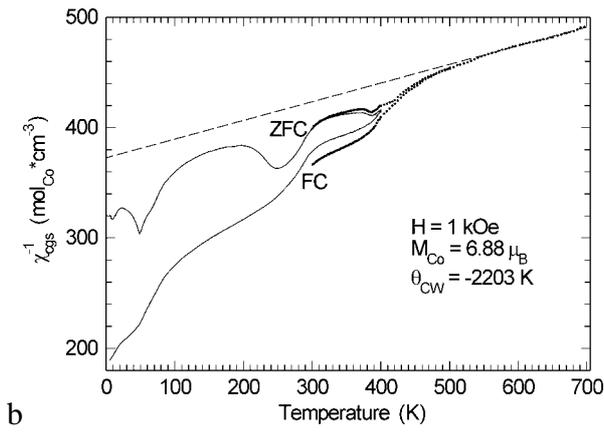

Fig.8

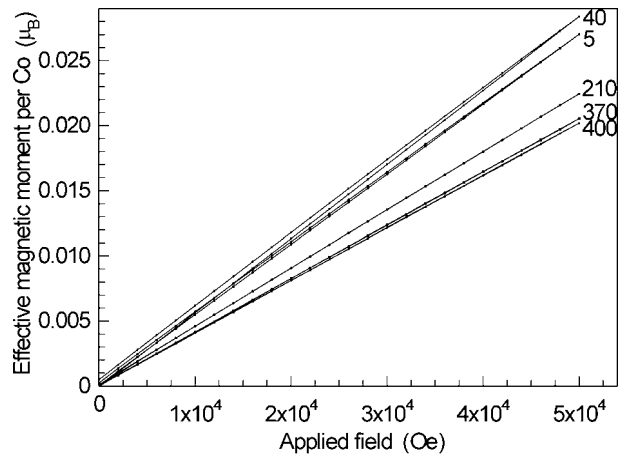

Fig.9

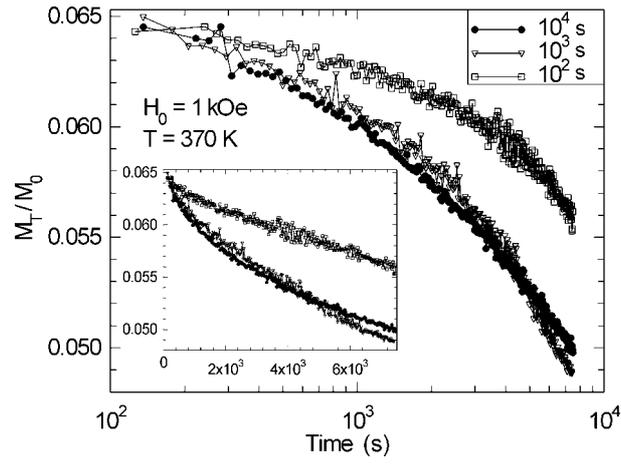

Fig.10

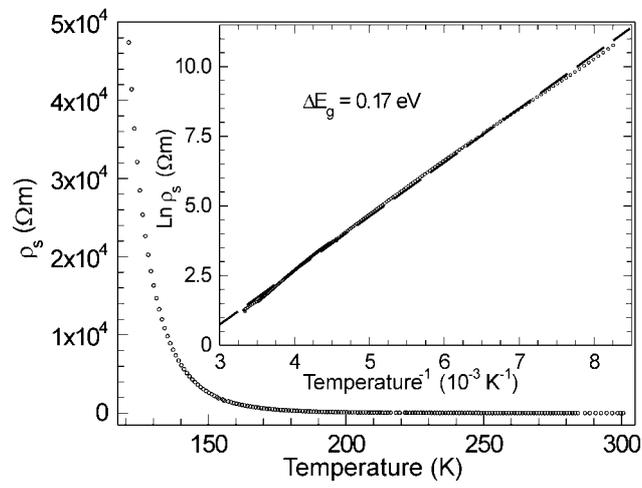

Fig.11